\documentclass{article}
\usepackage{newspr}
\usepackage{latexsym}
\newenvironment{eqnarraz}%
   {\setlength{\arraycolsep}{0.15em} \begin{eqnarray}}%
   {\end{eqnarray}}
\newcommand{\artit}[1]{#1}
\begin{document}
\title{THE COUNTERFACTUAL MEANING OF THE ABL RULE}
\author{L. MARCHILDON}
\address{D\'{e}partement de physique,
  Universit\'{e} du Qu\'{e}bec,\\
  Trois-Rivi\`{e}res, Qc.\ G9A~5H7, Canada\\
  E-mail: marchild@uqtr.ca}
\maketitle
\begin{abstract}
The Aharonov-Bergmann-Lebowitz rule assigns
probabilities to quantum measurement results at
time $t$ on the condition that the system is
prepared in a given way at $t_1 < t$ and found
in a given state at $t_2 > t$.  The question
whether the rule can also be applied
counterfactually to the case where no measurement
is performed at the intermediate time $t$ has
recently been the subject of controversy.
I argue that the
counterfactual meaning may be understood in terms
of the true value of an observable at $t$.
Such a value cannot be empirically determined for,
by stipulation, the measurement that would yield
it is not performed.  Nevertheless, it may or may
not be well-defined depending on one's proposed
interpretation of quantum mechanics.  Various
examples are discussed illustrating what can be
asserted at the intermediate time without
running into contradictions.
\end{abstract}
%
\section{Introduction}
The Aharonov-Bergmann-Lebowitz (ABL)
rule\cite{aharonov} was proposed in 1964
to compute measurement result probabilities
of systems that are both preselected and
postselected.  It was meant to provide
a time-reversal invariant formulation of
nonrelativistic quantum mechanics.

As long as a measurement is actually
carried out between pre- and postselection,
the ABL~rule is a straightforward consequence
of the Born probability rule and usual
assumptions on the quantum state following
a measurement.  Attempts have been made, however,
to interpret the ABL~rule counterfactually, that is,
to cases where no intermediate measurement is
made.  The resulting controversy is, I believe,
partly due to the fact that proposed definitions
of counterfactuals in the present
quantum-mechanical context do not adequately
capture the intuitive meaning of the notion.

In this paper I first review
the derivation of the ABL~rule and some
of its properties, in particular contextuality.
Next I analyze alleged proofs
of the impossibility of interpreting the rule
counterfactually, as well as various arguments
supporting or opposing them.  I attempt to clarify
the meaning of a counterfactual interpretation,
and point out circumstances in which the
counterfactual assertion of the ABL~rule
is or is not correct.
%
\section{The ABL~rule}
Consider a quantum system $S$ and three
observables $A$, $C$ and $B$ pertaining to it.
For simplicity, we shall first assume that these
observables are discrete and nondegenerate.

Suppose that at time $t_1$, $S$ is prepared in an
eigenstate $|a\rangle$ of $A$.  At $t > t_1$, the
observable $C$ is measured.  According to the Born
rule, the probability of obtaining result $c_i$
(where $\{c_j\}$ is the set of eigenvalues of $C$)
is given by $|\langle c_i |a\rangle|^2$.\footnote{All
kets are normalized.  We assume that the system's
Hamiltonian vanishes in between measurements.  If it
does not, evolution operators must be appropriately
inserted, or kets expressed in the Heisenberg
picture.}

Suppose the result $c_i$ is indeed obtained upon
measurement at $t$.  We assume the measurement 
interaction is such that immediately
after $t$, the system
is in state $|c_i \rangle$.  At a later time $t_2$,
the observable $B$ is measured.  The probability of
obtaining result $b$ is then given by
$|\langle b |c_i \rangle|^2$.  Hence the
probability of result $c_i$ and result
$b$, conditional on preparation $|a\rangle$,
is equal to
\begin{equation}
P(c_i \wedge b |a) = |\langle b |c_i \rangle
\langle c_i |a\rangle|^2 .
\label{joint}
\end{equation}

We are interested here in the probability of 
obtaining $c_i$ at $t$, on the condition that $S$ is
prepared in $|a\rangle$ at $t_1 < t$ and found
in $|b\rangle$ at $t_2 > t$.  This we shall denote by
$P(c_i|a,b)$.  From the definition of conditional
probability, we have (for $P(b|a) \neq 0$)
\begin{equation}
P(c_i |a, b) = \frac{P(c_i \wedge b |a)}{P(b|a)} . 
\end{equation}
Here $P(b|a)$ is the total probability of $b$
(given preparation $|a\rangle$), equal to the
sum of $P(c_j \wedge b |a)$ over all possible
results $c_j$.  From~(\ref{joint}) we obtain
\begin{equation}
P(c_i |a, b) 
= \frac{|\langle b |c_i \rangle \langle c_i |a\rangle|^2}
{\sum_j |\langle b |c_j \rangle \langle c_j |a\rangle|^2}
= \frac{\mbox{Tr} (P_b P_{c_i} P_a P_{c_i})}
{\sum_j \mbox{Tr} (P_b P_{c_j} P_a P_{c_j})} .
\label{abl}
\end{equation}
This is the ABL~rule\cite{aharonov}.  In the
right-hand side, $P_a$ and $P_b$ are projectors
on states $|a\rangle$ and $|b\rangle$.  The
operator $P_{c_i}$ projects on $|c_i \rangle$ or,
if the eigenvalue $c_i$ is degenerate, on the
associated subspace.\footnote{The sum over $j$
is then carried out on these subspaces.}
The ABL~rule (expressed in terms of projection
operators) also holds in that case, provided
the intermediate measurement obeys the 
L\"{u}ders rule $|a\rangle \rightarrow N
P_{c_i} |a\rangle$ (where $N$ is a normalization
constant).  The ABL~rule can also be written
for multiple intermediate measurements or for
selection by means of density matrices rather
than pure states, but we won't need these
generalizations here.

A much discussed property of the ABL~rule is
its contextuality.  Let $|u_1 \rangle$, 
$|u_2 \rangle$ and $|u_3 \rangle$ be an 
orthonormal basis in a three-dimensional Hilbert 
space.  Each of these kets can represent a state
wherein a particle is in one of three disjoint boxes.
Take the initial and final states as
{\begin{eqnarraz}
|a\rangle &=& {\textstyle\frac{1}{\sqrt{3}}}
\left\{ |u_1 \rangle + |u_2 \rangle
+ |u_3 \rangle \right\} , \label{ar} \\
|b\rangle &=& {\textstyle\frac{1}{\sqrt{3}}}
\left\{ |u_1 \rangle + |u_2 \rangle
- |u_3 \rangle \right\} . \label{br}
\end{eqnarraz}}%
Let the intermediate observable $C$ be so
chosen as to have $|u_1 \rangle$ as one of its
eigenvectors, corresponding to a nondegenerate
eigenvalue $c_1$.  The ABL~rule then yields
\begin{equation}
P(c_1|a, b) = \frac{\mbox{Tr} (P_b P_{c_1} P_a P_{c_1})}
{\sum_j \mbox{Tr} (P_b P_{c_j} P_a P_{c_j})} .
\end{equation}
It turns out that the right-hand side of this
equation is not well-defined unless $C$ itself
is well-defined.  Let $c_1$, $c_2$ and $c_3$ be
three different real numbers and define
\begin{equation}
C = c_1 |u_1 \rangle \langle u_1 |
+ c_2 |u_2 \rangle \langle u_2 | 
+ c_3 |u_3 \rangle \langle u_3 | .
\label{obsc}
\end{equation}
Elementary algebra shows that
$P(c_1|a, b) = 1/3$.  But if
\begin{equation}
C' = c_1 |u_1 \rangle \langle u_1 |
+ c_2 \{|u_2 \rangle \langle u_2 | 
+ |u_3 \rangle \langle u_3 | \} ,
\label{obsc1}
\end{equation}
then $P(c_1|a, b) = 1$.  Hence the ABL probability
of $c_1$ depends not only on the eigenspace 
associated with that eigenvalue, but also on the
structure of the observable in the orthogonal
eigenspace.  This is contextuality.  In terms of
boxes, this means the following.  If an observable
($C'$) distinguishes the first box from the
other two taken together, then the ABL probability
of being in the first box is~1.  If another
observable ($C$) distinguishes the
three boxes one from another, the ABL probability
of being in the first box is 1/3.

It is easy to check that if $C = A$,
\begin{equation}
P(a|a, b) = 1 . \label{pa}
\end{equation}
Likewise if $C = B$,
\begin{equation}
P(b|a, b) = 1 . \label{pb}
\end{equation}
From~(\ref{pa}) and~(\ref{pb}),
Albert, Aharonov and D'Amato have argued that
noncommuting observables $A$ and $B$ must
be simultaneously well-defined at 
$t$\cite{albert}.  This conclusion hinges
on a counterfactual use of the ABL~rule.  It was
indeed intended to apply specifically to the case
where neither $A$ nor $B$ are measured between
pre- and postselection.
%
\section{Counterfactual interpretation}
Can the ABL~rule be interpreted counterfactually?
That question was answered in the negative by
Sharp and Shanks\cite{sharp}, Cohen\cite{cohen}
and Miller\cite{miller}, and much debated afterwards.

The crux of the Sharp and Shanks argument
(as well as others) can be stated rather
succinctly.  Let $|b \rangle$ and $|b' \rangle$
be the possible final states of a two-state system.
Suppose $C$ is not measured, and assume that the
ABL~rule correctly gives the probability of 
nondegenerate result
$c_1$, conditional on pre- and postselection, had
$C$ been measured.  The total probability of $c_1$
should then be given as a weighted sum on the possible
final states, that is,
\begin{equation}
P(c_1 |a) = |\langle b|a \rangle|^2 P(c_1 |a,b)
+ |\langle b'|a \rangle|^2 P(c_1 |a,b') .
\label{ss}
\end{equation}
Here $|\langle b|a \rangle|^2$ is the probability
of final state $|b\rangle$ when no intermediate
measurement is made, and $P(c_1|a, b)$ is the
conditional probability of $c_1$ based on the
counterfactual interpretation of the ABL~rule.
But according to standard quantum mechanics,
$P(c_1 |a)$ is given by $|\langle c_1|a \rangle|^2$.
Hence we should have
\begin{equation}
|\langle c_1|a \rangle|^2
= |\langle b|a \rangle|^2 
\frac{\mbox{Tr} (P_b P_{c_1} P_a P_{c_1})}
{\sum_j \mbox{Tr} (P_b P_{c_j} P_a P_{c_j})}
+ |\langle b'|a \rangle|^2 
\frac{\mbox{Tr} (P_{b'} P_{c_1} P_a P_{c_1})}
{\sum_j \mbox{Tr} (P_{b'} P_{c_j} P_a P_{c_j})} .
\end{equation}
Since this is not true in general (counterexamples
are easily found), Sharp and Shanks
conclude that the counterfactual use is invalid.

The validity of the counterfactual use and the
relevance of the Sharp and Shanks argument were
debated between
Vaidman\cite{vaidman1,vaidman2,vaidman3,vaidman4,vaidman5}
and Kastner\cite{kastner1,kastner2,kastner3,kastner4}.
Vaidman's objection to the proof
consists in pointing out that 
the weight $|\langle b|a \rangle|^2$
in (\ref{ss}) is the probability of $b$ if no
intermediate measurement occurs.  Since we are asking
for the total probability of result $c_1$, we must use
the expression for the probability of $b$ if $C$ is
measured at $t$.  That probability is given by
$\sum_j \mbox{Tr} (P_b P_{c_j} P_a P_{c_j})$.
But then Eq.~(\ref{ss}) becomes
{\begin{eqnarraz}
P(c_1 |a)
&=& \sum_j \mbox{Tr} (P_b P_{c_j} P_a P_{c_j})
P(c_1 |a, b) + \sum_j \mbox{Tr} (P_{b'} P_{c_j} 
P_a P_{c_j}) P(c_1 |a, b') \nonumber \\
&=& \mbox{Tr} (P_b P_{c_1} P_a P_{c_1})
+ \mbox{Tr} (P_{b'} P_{c_1} P_a P_{c_1}) \nonumber \\
&=& \mbox{Tr} (P_{c_1} P_a P_{c_1} ) \nonumber \\
&=& |\langle c_1 |a \rangle |^2 ,
\label{ss1}
\end{eqnarraz}}%
in accordance with standard quantum mechanics.

The significance of this calculation of
$P(c_1 |a)$ is best brought out by quoting
Vaidman's definition of the
counterfactual meaning of the 
ABL~rule\cite{vaidman3}:
\begin{quote}
If a measurement of an observable $C$ were
performed at time $t$, then the probability for
$C = c_j$ would equal $P(c_j)$, provided that the
results of measurements performed on the system
at times $t_1$ and $t_2$ are fixed.
\end{quote}
This is a statement about the
statistical distribution of results of unperformed
experiments, made on the basis of a law
derived from a large number of performed experiments
\emph{identical to the former in all relevant
respects}.\footnote{Admittedly, our belief in the
noncounterfactual validity of the ABL~rule comes
from more indirect evidence, but this does not
affect the present discussion.}
It is like saying, on the basis of numerous
tosses of a die yielding essentially uniform
outcomes, that had the die been tossed one additional
time, the probability of obtaining ``3'' would have been
1/6.  Both this statement and Vaidman's definition
express the regularity of Nature.  Both assert, on the 
basis of a general rule drawn from numerous
experimental trials, that if the experiment were done
over again in the same relevant conditions, the results
would fall under the same general rule.  Hence
Vaidman's statement is (presumably) true, genuinely
counterfactual, but not particularly illuminating, at
least as far as specificities of quantum mechanics
are concerned.\footnote{The relation between
weak values and measurement results with unit ABL
probability\cite{vaidman4} does not affect that
remark, since it can be established
solely on the basis of the noncounterfactual use
of the ABL~rule.}

In contrast with Vaidman, Kastner defended the Sharp
and Shanks proof and argued that 
a nontrivial counterfactual assertion of the
ABL~rule (i.e.\ one that provides information
about specific systems) is false.  Kastner's
analysis draws from both Goodman's and Lewis's
theories of
counterfactuals\cite{goodman,lewis,horwich}.
Her argument can be summarized as follows.

A counterfactual $p \; \Box \hspace{-0.5em}
\rightarrow q$ (read ``If
it were that $p$, it would be that $q$.'') is true if
$p$ is not true and the conjunction of $p$ with
laws of nature $L$ and suitable background conditions
$S$ implies $q$, that is,
\begin{equation}
(p \wedge L \wedge S) \rightarrow q .
\label{pls}
\end{equation}
Now clearly, there must be restrictions
on $S$ for, if $S$ includes the statement $\neg p$
(``not $p$''), implication~(\ref{pls}) will hold
trivially.  These restrictions usually state that
$S$ should be ``cotenable'' with $p$, a notion not
so easy to define but essentially meaning that $S$ is
independent of the truth of $p$.

Suppose that in the real world, preparation
$|a\rangle$ is followed by postselection
$|b\rangle$, with no intermediate measurement.  Then
had there been an intermediate measurement,
the intermediate state would have changed,
and the final measurement result $b$ could no
longer be expected to obtain.  In other words,
result $b$ is not cotenable with measuring $C$.

Does it follow from this argument that the nontrivial
counterfactual assertion of the ABL~rule is false?
The answer is affirmative if the background
conditions must include the result $b$.  But this
is not necessarily the case.  The background 
conditions can be construed as encompassing everything
that characterizes the state of the system at $t$,
including whatever might induce it to yield result
$b$ upon measurement of $B$ at $t_2$.  With such a
definition, the result $b$ itself is not part of
the background conditions.  And although
cotenability might be analyzed in that context,
I shall use a different approach.

I submit that the proper way of investigating the
counterfactual validity of the ABL~rule is to
enquire about the ``true value'' of $C$ at $t$.
Strict empiricists will no doubt quit reading
right here, since it is absolutely impossible,
in a situation where $C$ has not been measured
between $t_1$ and $t_2$, to empirically ascertain
what the true value of $C$ was at the intermediate
time $t$.  But that doesn't prevent different
theories or interpretations to make claims
on what the true value is, as we will presently see.
%
\section{What can be said at $t$?}
It is instructive to analyze the Sharp and Shanks
argument in terms of true values.  If, following
standard quantum mechanics, $P(c_1 |a)$ is equal to
$|\langle c_ 1|a \rangle|^2$, then the first
equality in~(\ref{ss1}) holds identically.  But
Sharp and Shanks claim that the counterfactual
meaning of the ABL~rule is encapsulated in
Eq.~(\ref{ss}).  It is easy to see that (\ref{ss})
and~(\ref{ss1}) will coincide if
\begin{equation}
|\langle b|a \rangle |^2
= \sum_j \mbox{Tr} (P_b P_{c_j} P_a P_{c_j}) ,
\label{bcac}
\end{equation}
together with a similar equation with $b$ replaced
by $b'$.  In quantum mechanics, Eq.~(\ref{bcac})
does not hold in general.

The left-hand side of (\ref{bcac}) is the
uncontroversial probability of $b$ when no
intermediate measurement is performed, while
the right-hand side is the uncontroversial
probability of $b$ when $C$ is measured at the
intermediate time.  That the equality is
false means that whichever true value $C$ does
or does not have at the intermediate time $t$,
it cannot be one that is simply revealed in and
unaffected by an eventual measurement.

It was pointed out in\cite{cohen} that
(\ref{bcac}) holds if $(P_a, \{P_{c_j}\}, P_b)$ makes up a
consistent family of histories.\footnote{For
relevant definitions see\cite{marchildon},
Sect.~12.6.}  Indeed in that
case one can maintain that $C$ has a well-defined
value (equal to one of its eigenvalues) without
running into contradictions.  But of course one is
not compelled to do so.  Associating or not
associating true values to observables belonging to
a consistent family of histories contributes in
defining the interpretation of the theory.

It is instructive to recall the example of
Sect.~2, with $|a\rangle$ and $|b\rangle$ defined
as in Eqs.~(\ref{ar}) and~(\ref{br}).  Let $C$ and $C'$ be
defined as in~(\ref{obsc}) and~(\ref{obsc1}) and let
\begin{equation}
C'' = c_1 \{|u_1 \rangle \langle u_1 |
+ |u_3 \rangle \langle u_3 | \} 
+ c_2 |u_2 \rangle \langle u_2 | .
\label{obsc2}
\end{equation}
It is easy to check that
\begin{equation}
\left( P_a, \{P_{u_1}, I - P_{u_1}\}, P_b \right)
\label{h1}
\end{equation}
and
\begin{equation}
\left( P_a, \{P_{u_2}, I - P_{u_2}\}, P_b \right)
\label{h2}
\end{equation}
make up distinct consistent families of histories.
The ABL probability that $C' = c_1$ is one, and so is
the ABL probability that $C'' = c_2$.  We won't run
into contradictions if we maintain that the particle
was surely in box~1 at~$t$.  Likewise we won't run
into contradictions if we maintain that it was in
box~2.  Of course both statements cannot be held
at once, since they are logically contradictory.  Note
that the family of histories
\begin{equation}
\left( P_a, \{P_{u_1}, P_{u_2}, P_{u_3}\}, P_b \right) ,
\label{h3}
\end{equation}
more refined than either (\ref{h1}) or~(\ref{h2}),
is not consistent. 

The families $(P_a, \{P_{a_k}\}, P_b)$ and
$(P_a, \{P_{b_l}\}, P_b)$, where $\{a_k\}$ and $\{b_l\}$
stand for the set of eigenvalues of $A$ and $B$, are
always consistent.  Hence it can be maintained
that $A$ has the true value $a$ at intermediate times,
or that $B$ has the true value $b$.  We have assumed
that $P(b|a) \neq 0$.  Therefore $|b\rangle$ and $|a\rangle$
are not orthogonal.  So it is not a priori logically
inconsistent to assume that $A$ and $B$ both have true
values at intermediate times.\footnote{Note that
if $P_{ab}$ denotes the projector on the subspace
spanned by $|a\rangle$ and $|b\rangle$, then the
family of histories $(P_a, \{ P_{ab},
I - P_{ab}\}, P_b)$ is consistent.}  It is not
clear whether a full-fledged interpretation can
be implemented along these lines, for arbitrary
observables $A$ and $B$.

The ABL~rule is symmetric under permutation of
preselection and postselection.  This means that for
the purpose of making probabilistic statements about
intermediate measurement outcomes, the initial
and final states $|a\rangle$ and $|b\rangle$ have
exactly the same utility.  This does not necessarily
entail that they are equally useful for the purpose
of making ontological statements.  In von Neumann's
measurement theory\cite{neumann}, for instance,
a measurement is an interaction between a quantum
system and an apparatus, followed by a collapse.
From the time $t_1$ when the system is prepared
in state $|a\rangle$ to the time $t$ when an ontological
statement is to be made, no physical action occurs
on the system.  Such is not the case, however, between
$t$ and $t_2$.  Indeed a physical interaction of
the quantum system with an apparatus has to occur
sometime before $t_2$, for the system to collapse
to $|b\rangle$ at $t_2$.  In that context, it may be more
natural to hold that $|a\rangle$, rather than $|b\rangle$,
is the correct state at $t$, and that $A=a$, rather than
$B=b$, expresses a true value.

In some interpretations of quantum mechanics, true
values of observables can be assigned outside the
context of consistent families of histories.  Their
statistical distributions, however, will not obey
the ABL~rule.  An example is Bohmian mechanics,
where the true value of position is defined at any
intermediate time\cite{aharonov2}.  But in general,
$(P_a, \{P_x\}, P_b)$ does not make up a consistent
family.  In this context, the meaning of background
conditions proposed at the end of the last section
is particularly clear.  Suppose that $|a \rangle$
and $|b \rangle$ correspond to one-dimensional
wave functions $\psi_a (x)$ and $\psi_b (x)$,
with $\psi_a$ a Gaussian and $\psi_b$ a function
uniform over some interval $\Delta$ and zero
elsewhere (i.e.\ $|b \rangle$ postselects through
a slit of width $\Delta$).  All Bohmian trajectories
going through $\Delta$ at $t_2$ have gone through
some other interval $\Delta_t$ at~$t$.  Hence the
background conditions associated with preselection
$|a \rangle$ and postselection $|b \rangle$ are the
wave function at~$t$ together with the interval
$\Delta_t$ of true positions.  Of course if
position is measured at~$t$, the wave function
will change accordingly, and so will the measurement
result probabilities at~$t_2$.

It was pointed out by Vaidman\cite{vaidman3}
that no hidden variable theory can reproduce the
ABL~rule in all situations.\footnote{Vaidman is well 
aware that in hidden-variable theories, counterfactual
statements involve fixing the values of hidden
variables.  See for instance his discussion
in\cite{vaidman2}, Sect.~5.}
Indeed take an ensemble of spin 1/2 particles prepared
in the state $|a\rangle = |+; \hat{z} \rangle$.
If there is no intermediate measurement, postselection
in the state $|b\rangle = |a\rangle$ will in fact
introduce no additional selection.  Hence if no backward
causality is assumed, true values at $t$ must be the same
whether postselection does or does not occur, irrespective
of any hidden variables.  But then the probability
that a measurement at $t$ of the spin along $\hat{n}$
yields $+$ must be equal to
$\cos ^2 \{ \frac{1}{2} \cos ^{-1} (\hat{n} 
\cdot \hat{z}) \}$, which differs from the ABL~value.

In the Copenhagen interpretation of quantum
mechanics (or at least in the most
instrumentalistic versions of it),
an observable $C$ has a value only when
a measurement of $C$ indicates that value.
This is also the case in Mohrhoff's more recent
interpretation\cite{mohrhoff1,mohrhoff4},
which incorporates the ABL~rule explicitly.
The fact that the ABL~rule predicts a
statistical distribution of values of $C$ implies,
according to Mohrhoff, an objective fuzziness
of $C$ in the case where the experiment is not
performed.  I believe that statement is genuinely
and nontrivially counterfactual.\footnote{For a different
assessment see\cite{kastner4,kastner5}.}
It asserts that no unmeasured observable of no
individual system whatsoever has a true value in the
interval between pre- and postselection.
%
\section{Conclusion}
The noncounterfactual meaning of the ABL~rule
is not controversial.  The validity of the
rule is then a straightforward consequence of 
standard quantum mechanics and usual hypotheses on
the state of a quantum system immediately after
measurement.  The rule is also true counterfactually
if it simply expresses the reproducibility of
experiments.

A counterfactual meaning more relevant to the
specificities of quantum mechanics refers to the
true value of an observable at an intermediate
time, when no observable is actually measured
between pre- and postselection.  In general the
ABL~rule is then counterfactually false.  It can
be true, however, when associated with a
consistent family of histories or when asserting
objective fuzziness or nonvaluedness of observables
not being measured.

Although statements about true values between
pre- and postselection may not have definite
empirical meaning, they can fruitfully be viewed
as contributing to define the interpretation
of the quantum-mechanical theory.
%

%
\end{document}